\documentclass [12pt,a4paper]{article}
\usepackage{psfig}
\usepackage{amssymb}

\def\slantfrac#1#2{\hbox{$\,^{#1}\!/_{#2}$}}
\def\mdot{{\raisebox{1pt}{\hbox{$\stackrel{\bullet}{M}$}}}\ \!\!}
\def\apgt{\ {\raise-.5ex\hbox{$\buildrel>\over{\scriptstyle\sim}$}}\ }
\def\aplt{\ {\raise-.5ex\hbox{$\buildrel<\over{\scriptstyle\sim}$}}\ }

\def\q{\linebreak}

\begin{document}

\title{The Structure of Cool Accretion Disc\\ in
Semidetached Binaries}

\author{D.V.Bisikalo$^1$, A.A.Boyarchuk$^1$,\\
P.V.Kaygorodov$^1$, O.A.Kuznetsov$^{1,2}$ and T.Matsuda$^3$\\[5mm]
$^1$ Institute of Astronomy RAS, Moscow, Russia\\
$^2$ Keldysh Institute of Applied Mathematics, Moscow, Russia\\
$^3$ Department of Earth and Planetary Sciences, Kobe
University,\\ Kobe, Japan\\}

\date{}
\maketitle

\begin{abstract}

We present the results of qualitative consideration of possible
\q changes occurring during the transition from the hot accretion
disc to the cool one. We argue the possible existence of one more type
of spiral density waves in the inner part of the disc where
gasdynamical perturbations are negligible. The mechanism of
formation of such a wave as well as its parameters are considered.

We also present the results of 3D gasdynamical simulation of cool
accretion discs. These results confirm the hypothesis of possible
formation of the spiral wave of a new, ``precessional" type in the
inner regions of the disc. Possible observational manifestations
of this wave are discussed.

\end{abstract}

\section{Introduction}

The analysis of principal processes of matter heating and cooling
in accretion discs presented in the work [\ref{di12}] has shown
that for realistic parameters of accretion discs in semidetached
binaries ($\mdot\simeq10^{-12}\div10^{-7}M_\odot/\mbox{year}$ and
$\alpha\simeq10^{-1}\div10^{-2}$)\footnote{$\mdot$ -- mass
transfer rate; $\alpha$ -- dimensionless parameter introduced by
Shakura and Sunyaev [\ref{shakura}, \ref{shak2}] for expression of
viscosity coefficient $\nu=\alpha c_s H$ ($H$ -- disc
semithickness, $c_s$ -- sound speed).} the gas temperature in
outer part of the disc lies in the range from $\sim 10^4$~K to
$\sim 10^6$~K. Earlier we have conducted 3D gasdynamical
simulation of accretion discs both for a `hot' case (the gas
temperature in the outer part of the disc was 200--500 thousands~K
[\ref{di1}--\ref{di11}]), and for the `cool' case (the temperature
of gas in the outer part of the disc didn't exceed $\sim 2 \times
10^4$~K [\ref{di12}]). The analysis of these results has shown
that for both cases (i.e. independently on the disc temperature)
the self-consistent solution didn't involve the shock interaction
between the stream of matter from the inner Lagrangian point $L_1$
and formed accretion disc (``hot spot"). Energy release zone
(``hot line") is located outside the disc and is formed as the
result of the interaction of the circumdisc halo and circumbinary
envelope with the stream.  The ``hot line" model is found to be in
a good agreement with observations
[\ref{Tanya98}--\ref{Tanya2003b}].


For `hot' solutions we have investigated both general morphology
of gaseous flows in semidetached binaries and the structure of
formed hot accretion discs (see, e.g., [\ref{di11},\ref{ippeg}]).
In particular, we have found only one arm of spiral shock wave
generated by the tidal influence of the mass-losing star. The
two-armed spiral shocks were discovered by Matsuda {\it et al.} in
[\ref{Sawada86}--\ref{spiral2}]. Nevertheless, our 3D gasdynamical
simulations for the `hot' case have shown the presence of only
one-armed spiral shock while in the place where the second arm should
be the stream from $L_1$ dominates and presumably prevents the formation of
the second arm of tidally induced spiral shock. Besides we have
found out that for the `hot' case the variation of mass transfer
rate leads to the disc perturbations and to the formation of
spiral-vortex structure in the disc
[\ref{blob1}--\ref{fridman}].


Even a glance at the morphology of gaseous flows for the
`cool' case (see, e.g., [\ref{di12},\ref{rujichka}]) discovers
that the accretion disc for this case is characterized by
principally different parameters as compared to the `hot' case. In
particular, the disc has more circular form and the second arm of
the tidal spiral wave is present. Our analysis [\ref{di12}] shows
that opposite to the `hot' case tidal spiral waves don't propagate
to the inner part of the cool accretion disc and are located in the
outer part of the disc only.

The aim of this work is the investigation of the structure of cool
accretion discs in semidetached binaries. Section~2 contains the
qualitative analysis of changes occurring in transition from a hot
accretion disc to the cool one. In particular, here we suggest the
possible existence of one more type of spiral density waves in the
inner part of the disc where gasdynamical perturbations are negligible.
We also consider the mechanism of formation of such a wave and its
parameters. Section~3 contains the results of 3D gasdynamical
simulation of flow structure for the case when radiative cooling
is effective and the gas temperature drops down to $\sim 10^4$~K in
the whole region. These results confirm the hypothesis of
possible formation of a spiral wave of a new type in the inner
regions of the disc.  Possible observational manifestations of
the discovered spiral wave of the new, ``precessional" type as well
as main conclusions~are~drawn~in~Section~4.

\renewcommand{\thefigure}{1}
\begin{figure}[t]
\centerline{\hbox{\psfig{figure=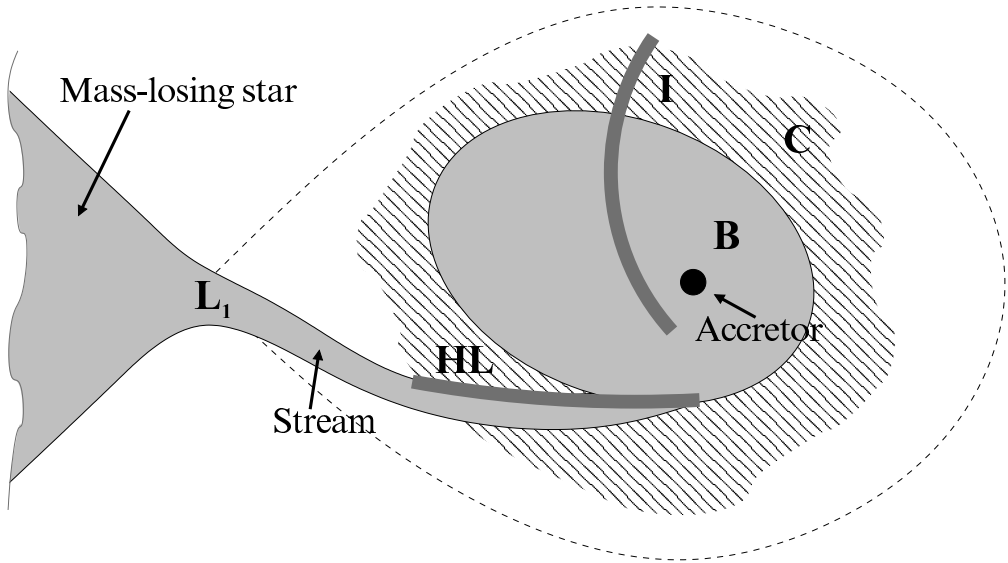,width=\textwidth}}}
\caption{\small
The sketch of main peculiarities of the morphology of gaseous
flows in semidetached binaries for the case of high gas
temperature.}
\end{figure}

\renewcommand{\thefigure}{2}
\begin{figure}[t]
\centerline{\hbox{\psfig{figure=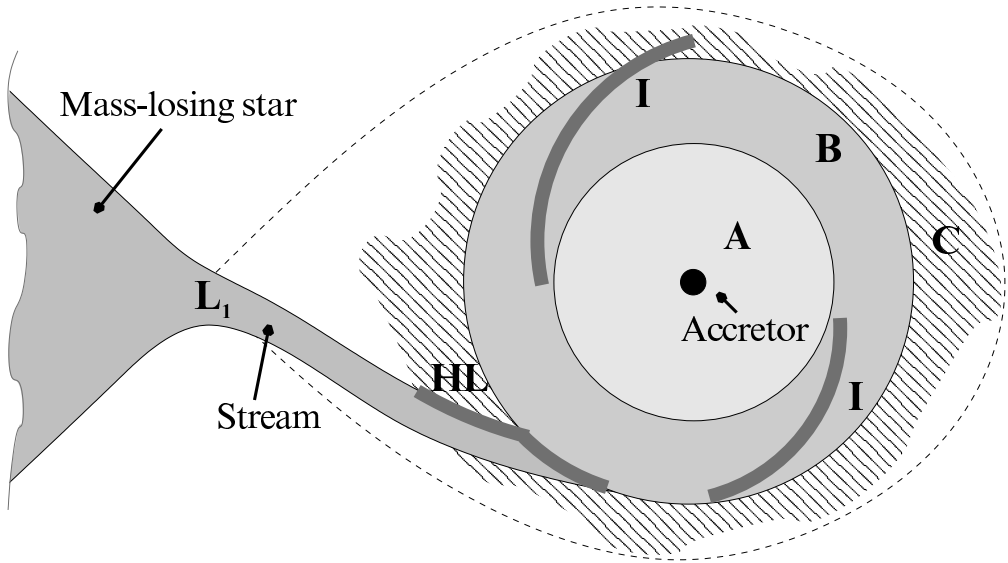,width=\textwidth}}}
\caption{\small
The sketch of main peculiarities of the morphology of gaseous
flows in semidetached binaries for the case of low gas
temperature.}
\end{figure}

\section{The structure of the cool accretion disc.\\  Qualitative
consideration.}

The sketch of main peculiarities of the morphology of gaseous
flows in semidetached binaries for the `hot' case is given in
Fig.~1. This scheme is based on the results of 3D gasdynamical
simulations published in [\ref{di1}--\ref{di11}]. In Fig.~1 the
fragment of mass-losing star that fills its Roche lobe, the
location of the inner Lagrangian point $L_1$, the stream of
matter from $L_1$, as well as the location of the accretor are
shown. A dashed line marks the Roche lobe. Following the
definitions given in [\ref{di8}] the morphology of gaseous flows
in semidetached binaries is governed by the stream of matter
from $L_1$, quasi-elliptical accretion disc, circumdisc halo and
circumbinary envelope. This classification of the main
constituents is based on their physical properties: (i) if the
motion of a gas particle is not determined by the gravitational
field of accretor then this particle belongs to the circumbinary
envelope filling the space between the components of binary;
(ii) if a gas particle revolves around the accretor and after that
mixes with the matter of the stream then it doesn't belong to
the accretion disc but forms the circumdisc halo (zone `C' in
Fig.~1); (iii) the accretion disc is formed by the matter of the
stream which is gravitationally captured by the accretor and
hereinafter doesn't interact with the stream but moves to the
accretor losing the angular momentum (zone `B' in Fig.~1). The
interaction of matter of circumdisc halo and circumbinary
envelope with the stream results in the formation of the shock
located along the edge of the stream. This shock is referred as
``hot line" and is marked by `HL' in Fig.~1. Tidal action of
mass-losing star results in formation of spiral shock marked by
`I' in Fig.~1.  Our 3D gasdynamical simulations for the `hot'
case have shown only one-armed spiral shock while in the place
where the second arm should be the flow structure is determined
by the stream from $L_1$. It also should be stressed that the
spiral shock deeply penetrates to the inner part of the disc in
this case.

Let us consider the changes occurring during the transition from
the hot accretion disc to the cool one. The sketch of main peculiarities
of the morphology of gaseous flows in semidetached binaries for
the `cool' case when non-adiabatic processes of radiative heating
and cooling result in dropping of gas temperature is given in
Fig.~2. Our 3D gasdynamical simulations presented in [\ref{di12}]
have shown that for the `cool' case when the radiative cooling
decreases the gas temperature to $\sim 10^4$~K the solution has
the same qualitative features as that for the `hot' case,
namely: the interaction between the stream and the disc is
shockless, the energy release zone -- shock wave `HL' -- is due to
the interaction with circumdisc halo and located outside the disc,
being rather elongated this shock wave can also be referred as
``hot line". At the same time, in the `cool' case the accretion
disc (zones `A' and `B' in Fig.~2) is significantly more dense as
compared to the matter of the stream, the disc is thinner and has
not quasi-elliptical but circular form. The size of circumdisc
halo is less as well.  The second arm of the tidal spiral shock is
formed, the both arms don't reach the accretor but are located in
the outer part of the disc. Taking into account that the stream
acts on the dense inner part of the disc weakly as well as that
all the shocks (``hot line" and two arms of tidal wave) are located in
the outer part of the disc we can introduce a new element of flow
structure for the `cool' case: the inner region of accretion disc
(zone `A' in Fig.~2) where the influence of gasdynamical
perturbations mentioned above is negligible.

Let us consider the flow of matter in the inner parts of the disc
that are not subjected by gasdynamical perturbations.
In the absence of external action a gas particle should
revolve around the gravitation center (accretor) along the
elliptical orbit. In our gasdynamical solutions (see Section~3)
gas particles move along the near-circular but elliptical
orbits, the accretor being located in one of the ellipse
focal points. It is known (see, e.g., [\ref{kumar}, \ref{warner}]),
that the influence of companion star results in retrograde
precession of particle's orbit, the precession rate is
proportional to orbit radius in accordance to \begin{equation}
\frac{P_{pr}}{P_{orb}} \simeq \slantfrac{4}{3}
\frac{\sqrt{1+q}}{q} \left( \frac{r}{A} \right)^{-3/2} \,,
\label{eq1} \end{equation} where $P_{pr}$ -- period of orbit
precession; $P_{orb}$ -- orbital period of binary; $q$ --
components' mass ratio; $r$ -- orbit radius; $A$ -- binary
separation.

The accretion disc is formed by a multitude of particles, each
of them moving along its own elliptical orbit. Due to interaction
between particles the disc should be considered in gasdynamical
approximation, so we should consider flowlines instead of orbits,
the first ones are also being elliptical. It is known that
flowlines can't intersect and can only be tangent to each other.
It is also evident from geometrical consideration that in order to
build the disc from non-intersecting ellipses we should embed it
into each other. For the case of zero eccentricity we will obtain
the circular disc.  For the case of non-zero eccentricity of
flowlines we can build the equilibrium solution with aligned
semimajor axes of ellipses.  In the presence of an external action
(as it occurs in binaries) orbits begin to precess, the precession
of flowlines more distant from the accretor being faster so it
will overtake the flowlines with less semimajor axes. Due to
impossibility of intersections between flowlines, the solution
goes to equilibrium state and all flowlines tend to precess with
the same angular velocity, i.e. in rigid-body mode. In this state
a remote orbit should be turned at larger angle in the direction
opposite to that of the matter rotation in the disc since the
precession is retrograde. The precession rate value lies in the
range from the one for outer (`fast') orbits and the precession
rate for inner (`slow') orbits. The inner orbits are defined as
those with the negligible gravitational influence of companion
star as compared to the gravitational influence of accretor.  The
outer orbits are defined as those lying in the region without
gasdynamical perturbations since latter can violate the regularity
of precession. The location of both innermost and outermost orbits
obviously depends on parameters of binary and parameters of mass
transfer (e.g., mass transfer rate) so we can expect the various
values of precession rate for different systems.

\renewcommand{\thefigure}{3}
\begin{figure}[t!]
\centerline{\hbox{\psfig{figure=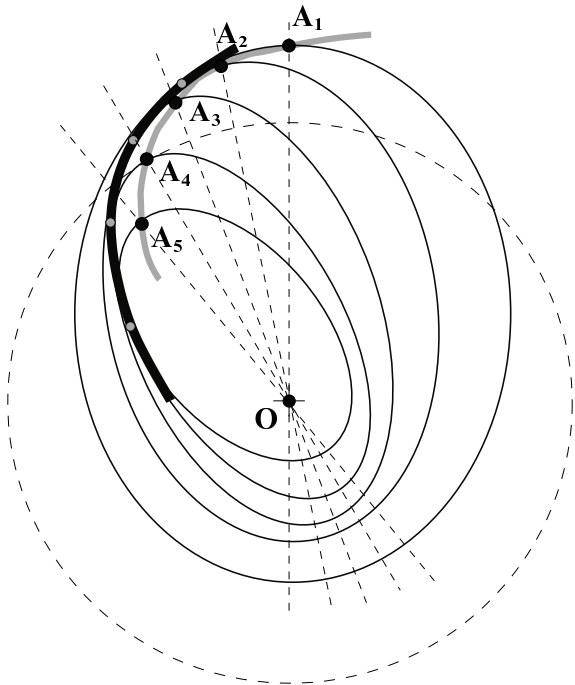,width=0.55\textwidth}}}
\caption{\small
The scheme of generation of spiral structures in the cool disc's
inner part where gasdynamical perturbations are
negligible. Apastrons for each flowline are shown by black
circles and marked by $A_1$, $A_2$, ..., $A_5$. The places of
maximal attachment of flowlines are shown by gray circles.}
\end{figure}

Let us consider a solution with the semimajor axis misaligned
w.r.t. some chosen direction and the angle of misalignment (turn
angle) being proportional to the value of semimajor axis (Fig.~3).
It is evident that such a solution should contain spiral
structures. In particular, the non-uniformity of the motion along
the elliptical flowline will result in increasing of density in
apastron, so the curve passing through apastrons (black circles
marked by $A_1$, $A_2$, ..., $A_5$ in Fig.~3) will be a spiral
density wave. The curve connected places of maximal attachment of
flowlines (gray circles in Fig.~3) is the spiral density wave as
well. After passing the apastron the particle's velocity
increases (including the radial component of velocity $v_r$) so we
can expect the increasing of radial component of matter flux
$F_{rad} \propto \rho v_r$ since both $\rho$ and $v_r$ increase.
Note, if our mechanism is consistent, the gas particle moving
along the flowline should show increase of density first and after
that the increase of radial velocity. Hence, the density wave
should precede the pike of radial component of flux matter, and
the curve passing through these peaks also will be a spiral. The
increasing of radial component of flux matter after passing the
wave will result in accretion rate increase in the region where
``precessional" wave touches the accretor.

Resuming the qualitative consideration of changes occurring during the
transition from a hot accretion disc to the cool one we can state
the following: (i) the region that is not subjected by
gasdynamical perturbations appears in the cool accretion disc;
(ii) the retrograde precession generates the spiral density wave
in the disc's inner parts that are not subjected by
gasdynamical perturbations; (iii) the rotational velocity of
this wave is determined by the mean precession rate of
flowlines; iv) the increase of radial component of the flux of matter
takes place after gas particles pass the wave, and the curve
connecting the peaks of radial flux of matter has a spiral shape.

\section{3D gasdynamical simulation of cool\\
accretion discs. Model and results.}

In order to perform the further analysis of possible existence of
a spiral wave of ``precessional" type in inner regions of cool
accretion discs we have conducted 3D gasdynamical simulation of
the disc structure for the case when the radiative cooling decreases
the gas temperature to $\sim 10^4$~K. We used rather fine
difference grid ($121\times121\times32$ gridpoints along $X$,
$Y$, $Z$ axes, correspondingly) to achieve good spatial
resolution in the inner part of the disc. We parallelized our
code and used supercomputer of Moscow Supercomputer Center, so
we could obtain solutions with good resolution on the timescale
greater than period of disc precession.

Details on our numerical model can be found in [\ref{di12}]. We
have taken the binary with parameters of dwarf nova IP Peg and
adopted the mass of accretor as $M_1=1.02M_\odot$, the mass of
mass-losing star as $M_2=0.5M_\odot$, and binary separation as
$A=1.42R_\odot$.  The finite-difference Roe-Osher method
[\ref{roe},\ref{osher}] was used to solve the gasdynamical
equations. This method was tuned for solving on multiprocessor
computers. 2D decomposition of computational grid with
synchronization of boundary conditions was used, so one processor
operated with a ``stick" of cells [\ref{prep}].  We conducted our
simulations in corotating non-inertial Cartesian coordinate frame
in the upper semi-space (due to symmetry of the problem w.r.t. the
equatorial plane). We have imposed free boundary conditions on
outer boundaries: constant density $\rho=10^{-8}\rho_{L_1}$
($\rho_{L_1}$ -- density of matter in $L_1$ point), temperature
13600~K, and zero velocity. The stream was also specified in the
form of a boundary condition: matter with specified temperature,
density, and velocity was injected into a zone around $L_1$ with
radius $0.014A$. Adopted parameters give the mass transfer rate
equal to $\sim 10^{-9} M_{\odot}/\mbox{year}$. The accretor was
adopted as a sphere of radius $10^{-2} A$. All matter that comes
to any of cells forming the accretor was taken to fall onto the
star. Computational domain was given in such a way that the disc
and the stream of matter from $L_1$ would be inside the
computational domain. We adopt the solution for the model without
cooling [\ref{ippeg}] as initial condition. The run of model with
cooling was conducted up to the time $\approx 10$ binary's period,
the computational time on supercomputer MVS1000A of Moscow
Supercomputer Center was approximately 2000 hours.

\renewcommand{\thefigure}{4}
\begin{figure}[p]
\begin{center}
\hbox{\hspace*{2cm}\psfig{file=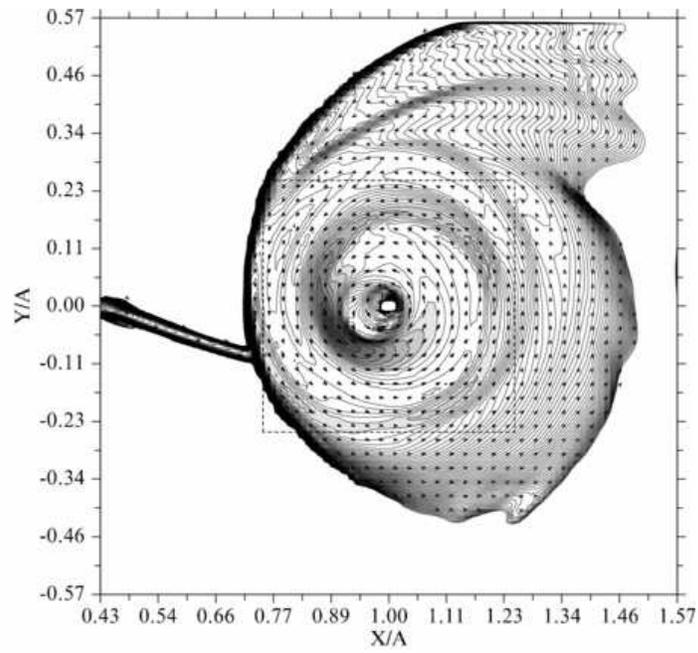,height=9cm}}
\hbox{\hspace*{2cm}\psfig{file=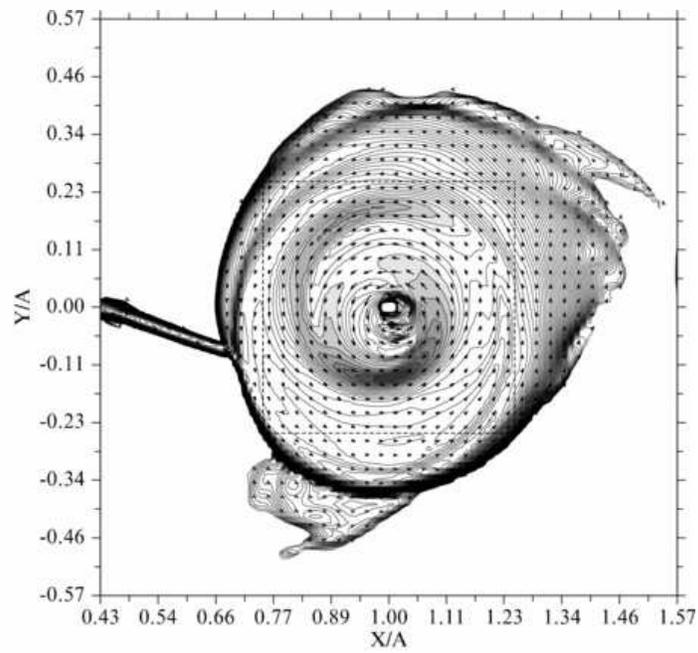,height=9cm}}
\end{center}
\caption{\small
Density isolines and velocity vectors in the equatorial plane of
binary for two moments of time $t=1.26P_{orb}$ and
$t=2.82P_{orb}$.}
\end{figure}

\renewcommand{\thefigure}{5}
\begin{figure}[p]
\begin{center}
\hbox{\hspace*{2cm}\psfig{file=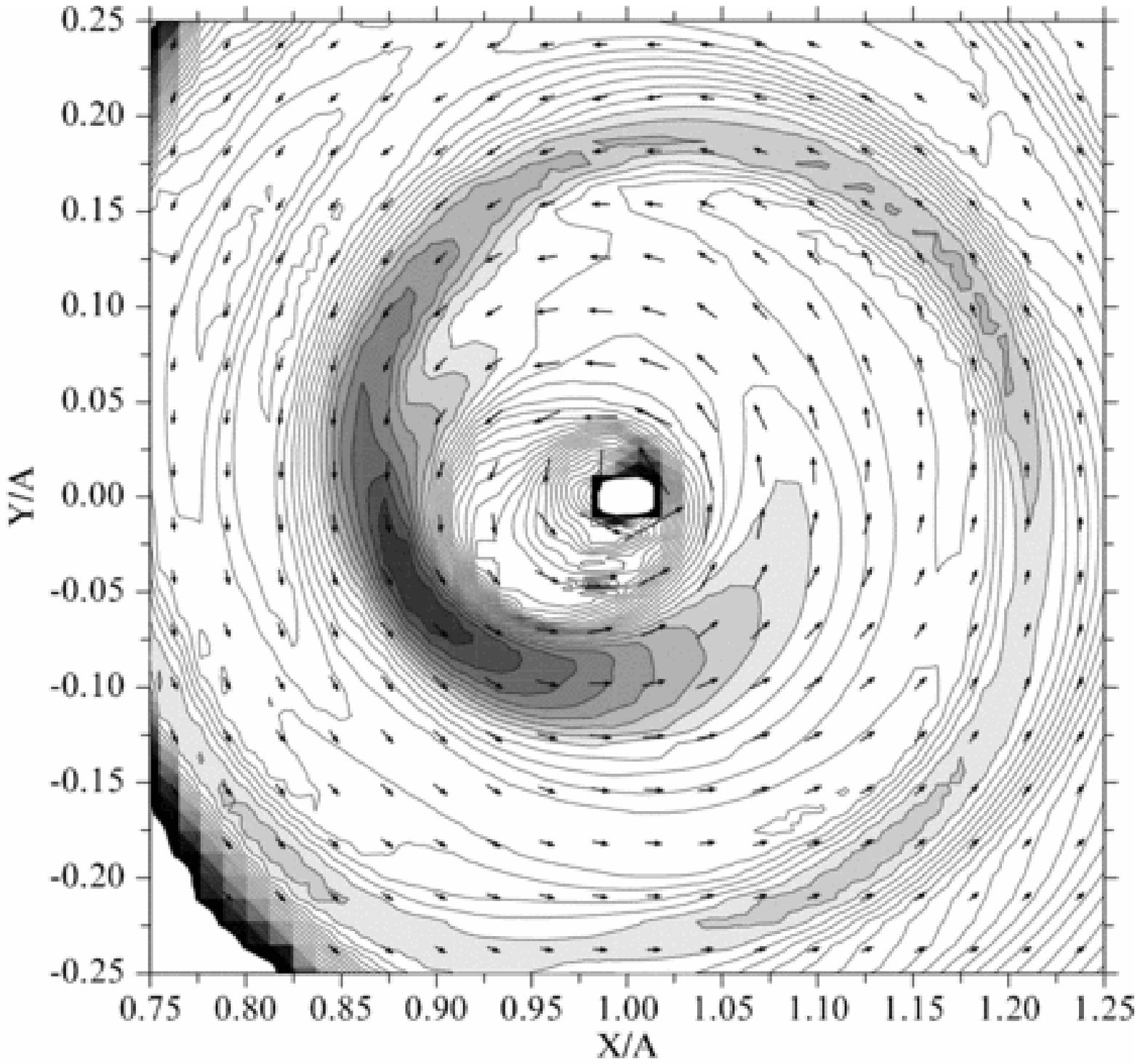,height=9cm}}
\hbox{\hspace*{2cm}\psfig{file=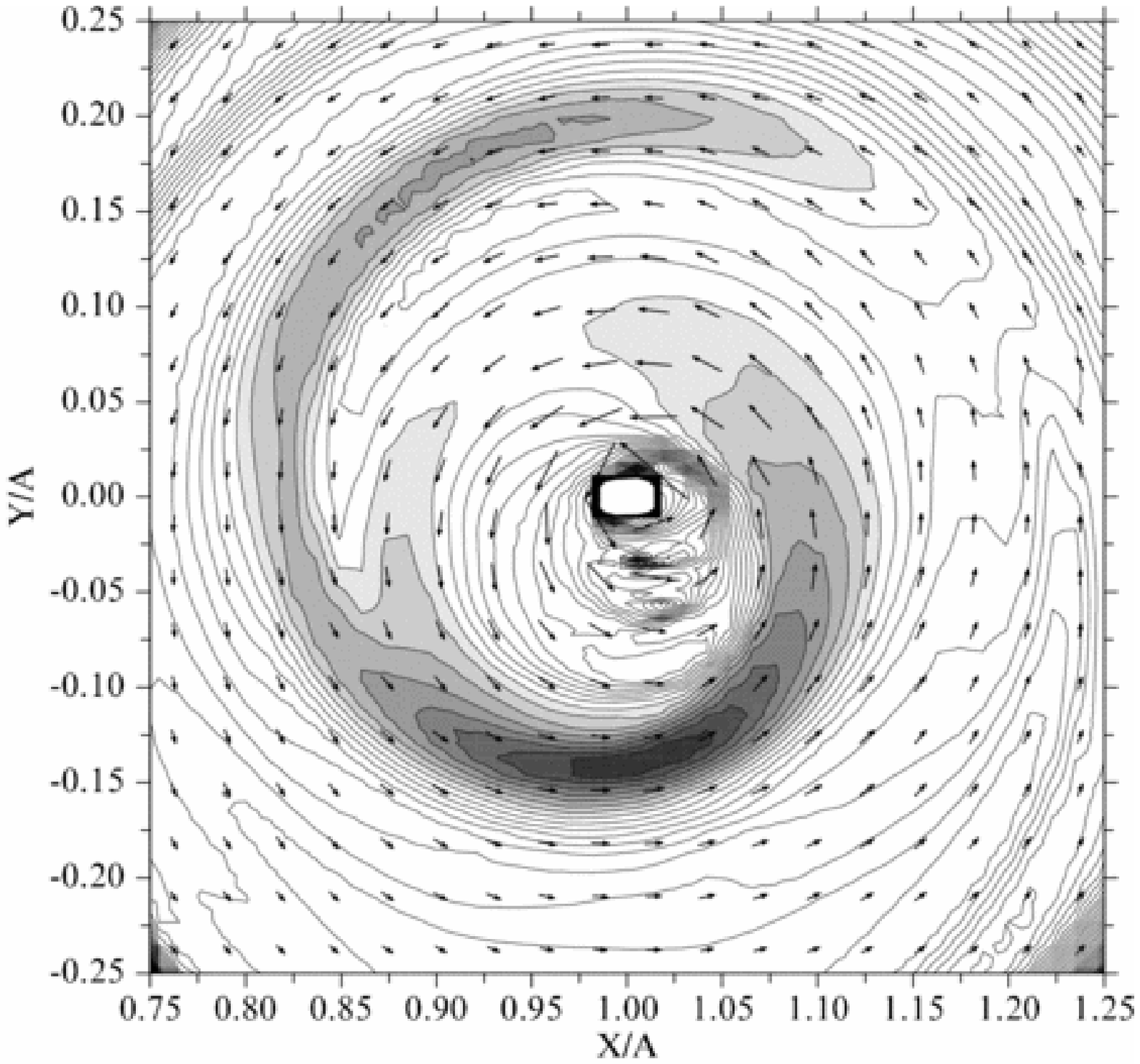,height=9cm}}
\end{center}
\caption{\small
Density isolines and velocity vectors in the inner part of the
disc (a dashed rectangular on Fig.~4) for the same moments of
time as in Fig.~4.}
\end{figure}

The morphology of gaseous flows in considered binary is shown in
Fig.~4 and Fig.~5. Figure~4 depicts the density distribution and
velocity vectors in the equatorial plane of the system for two
moments of time $t=1.26 P_{orb}$ and $t=2.82 P_{orb}$. The shocks,
which are formed in the disc, are seen as condensed isolines. The
latter on the edge of the circumdisc halo correspond to sharp
decrease of density up to the background value. It is seen that
the calculated flow structure is similar to that on the sketch
given in Fig.~2. We can see the dense circular disc as well as the
compact circumdisc halo. The interaction of gas of the circumdisc
halo with the stream generates the shock -- ``hot line", the
latter being located outside the disc. The two-armed spiral shock
wave is formed in the disc. The both arms are located in the
outer part of the disc. We can also see one more spiral wave
located in gasdynamically unperturbed region (see scheme in
Fig.~2). Figure~5 shows the zoom of density distribution and
velocity vectors in the inner part of the disc for the same
moments of time as in Fig.~4.  Figures~4 and~5 are presented in
corotational coordinate frame (i.e. in the frame rotating with
the orbital period of binary). It is seen that the two-armed
spiral wave is in rest for this coordinate frame (that is rather
natural for a tidal wave) but the inner spiral wave rotates. The
analysis of computational results shows that the wave moves as a
single whole and its velocity in the inertial frame (i.e. in the
observer's frame) turns out to be equal to $\approx -0.133$
revolution per one orbital period of the binary. In other words,
the spiral wave commits a full revolution in $\approx 7.5$ of
binary period (in inertial frame), i.e.  $P_{pr} \simeq 7.5
P_{orb}$. The wave terminates at the distance of $\approx
0.16 A$ from the accretor. The theoretical value of period in
accordance to (1) is equal to $\approx 16 P_{orb}$, i.e.
coincides with the value computed in gasdynamical model within
the accuracy of factor of two.

\renewcommand{\thefigure}{6}
\begin{figure}[t]
\centerline{\hbox{\psfig{figure=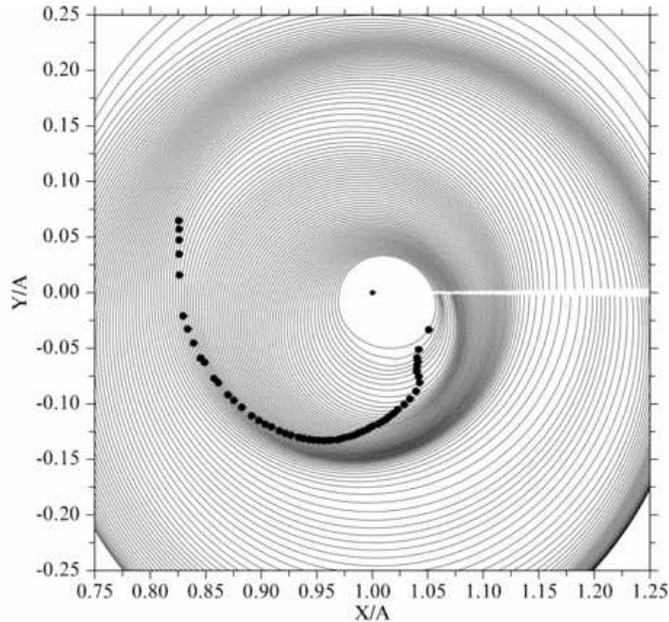,height=9cm}}}
\caption{\small
Calculated flowlines in the inner part of the disc for the
moment of time $t=2.82 P_{orb}$. Apastrons for each flowline
are shown by black circles.}
\end{figure}

\renewcommand{\thefigure}{7}
\begin{figure}[t]
\vspace*{-1cm}
\centerline{\hbox{\psfig{figure=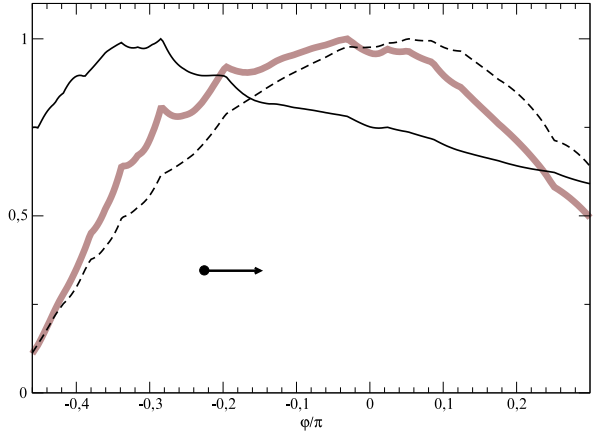,width=0.7\textwidth}}}
\caption{\small
Distribution of density (a solid line), radial velocity (a
dashed line) and radial flux of matter (bold gray line) along
some flowline in the vicinity of its apastron for the same
moment of time as in Fig.~6. An arrow shows the direction of
motion. All distribution are normalized on its maximum values.}
\end{figure}

\renewcommand{\thefigure}{8}
\begin{figure}[p]
\begin{center}
\vspace*{-5mm}
\hbox{\hspace*{-2cm}\psfig{file=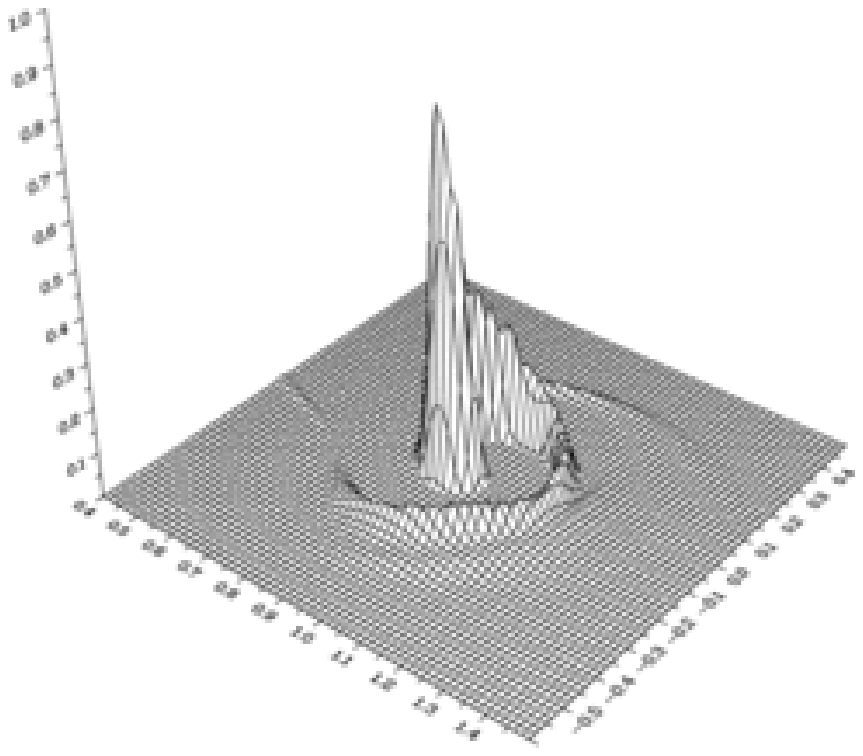,height=12cm}}
\vspace*{-2cm}
\hbox{\hspace*{2cm}\psfig{file=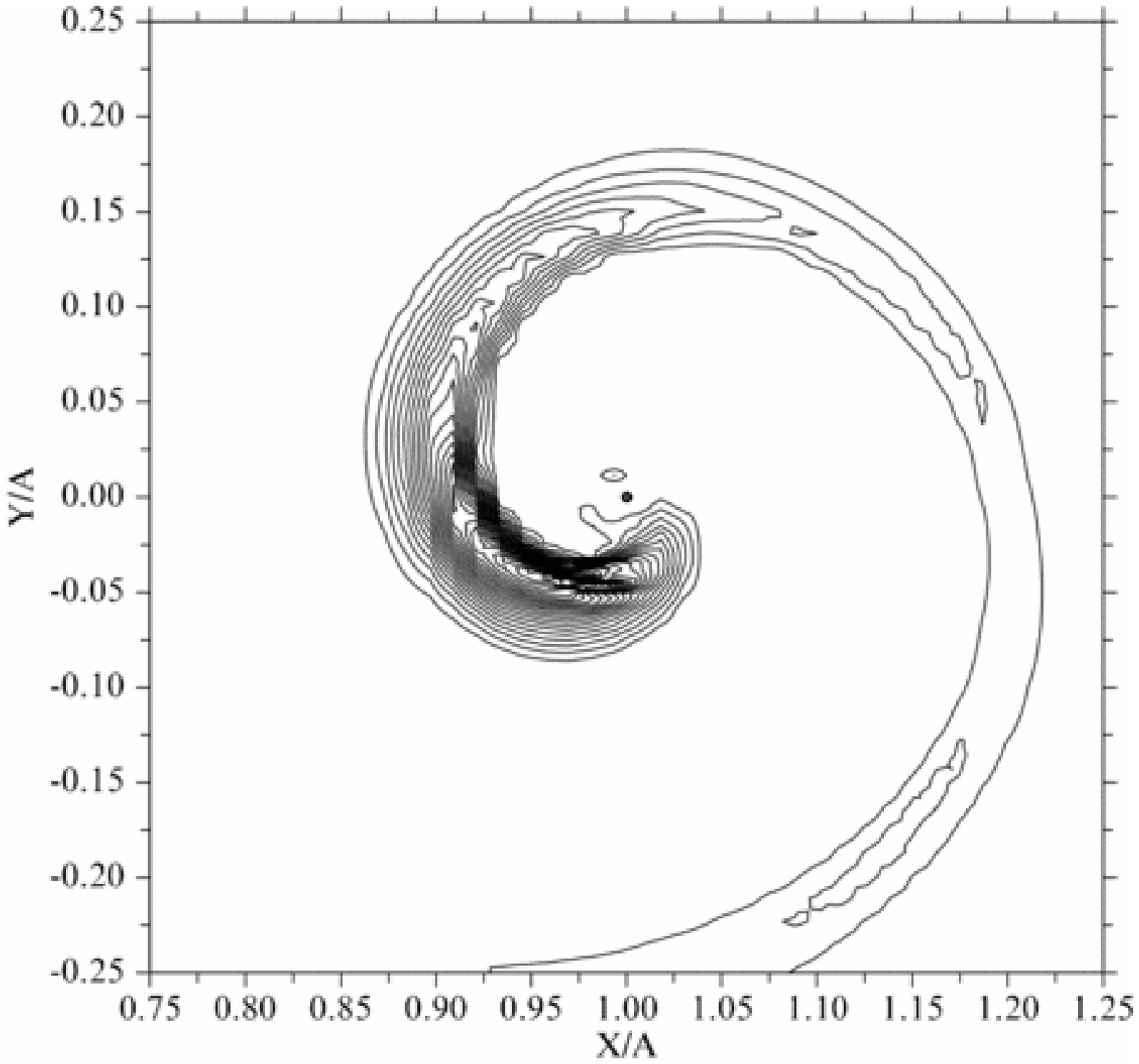,height=9cm}}
\vspace*{-0.5cm}
\end{center}
\caption{\small
Distribution of radial flux of matter in the equatorial plane
for the moment of time $t=4.92P_{orb}$. The flux is normalized
on its maximum values.}
\end{figure}

If our hypothesis on the precessional mechanism of generation
of inner spiral wave is consistent then computed flowlines
should behave in accordance with Fig.~3. The computed ones as
well as the apastrons for each flowline are shown in Fig.~6.
Note that the form and the position of the curve passing through
apastrons coincides with those ones for density wave drawn in
Fig.~5 for the same moment of time. So the computed structure is
in a good agreement with our qualitative consideration. Indeed,
the curve passing through apastrons constitutes the spiral
density wave, the places of maximum attachment of flowlines
being detached from this wave and having a spiral form as
well. The distributions of density, radial velocity and radial
flux\footnote{per unit of square.} of matter along the flowline
also correspond to our qualitative consideration (Fig.~7, the
portion near the apastron of the flowline is shown). When
moving along the flowline the density increases first (after
passing the apastron) and after that the radial velocity
increases, so in accordance to our forecast, the density wave precedes the pike of radial
component of flux matter. Figure~8
depicts the distribution of the radial flux of matter in the
equatorial plane of the disc for the time $t=4.92 P_{orb}$ in
two forms: bird-eye view and contours (the flux was normalized
to its maximal value). We can see distinctly the spiralwise form
of the curve passing through the flux peaks. The
accretion rate increases up to the order of magnitude in
comparison to the wave-free solution due to the increasing of radial
flux of matter behind the ``precessional" density wave.

\section{Conclusions}

The qualitative analysis of possible changes occurring in
transition from a hot accretion disc to the cool one shows the
possible generation of spiral wave of a new type in the
accretion disc. The appearance of this spiral wave in the inner part
of the disc where gasdynamical perturbations are negligible is
due to retrograde precession of flowlines in the binary system.

The analysis of presented results of 3D gasdynamical simulation
fully confirms our hypothesis on the possible generation
of spiral wave in the inner part of the cool disc. The
correspondence between the qualitative analysis and the
 computational results permits us to argue the
precessional mechanism of the wave formation.
Increasing of the radial flux of matter after passing the density
wave results in growth of accretion rate and formation of a compact
zone of energy release on the accretor surface.
This zone can be seen as a periodic increase of brightness on
light curves of semidetached binaries. Observation of this zone
will permit to determine the precession rate of the wave so
these observations will both give the proof of the existence of
``precessional" wave in the inner part of the cool accretion disc
and provide the information on characteristics of the inner
parts of the disc.

\section*{Acknowledgements}

The work was partially supported by Russian Foundation for Basic
Research (projects NN 02-02-16088, 02-02-17642, 03-01-00311,
02-02-16622), by Science Schools Support Program (project N
162.2003.2), by Federal Programme "Astronomy", by Presidium RAS
Programs "Mathematical modelling and intellectual systems",
"Nonstationary phenomena in astronomy", and by INTAS (grant N
00-491).  OAK thanks Russian Science Support Foundation for
the financial support.

\section*{References}

\begin{enumerate}

\item
\label{di12} D.V.Bisikalo, A.A.Boyarchuk, P.V.Kaygorodov and
O.A.Kuznetsov, Astron. Reports {\bf 47}, 809 (2003).

\item
\label{shakura} N.I.Shakura, Sov. Astron. {\bf 16}, 756 (1972).

\item
\label{shak2} N.I.Shakura and R.A.Sunyaev, Astron. Astrophys.
{\bf 24}, 337 (1973).

\item
\label{di1} D.V.Bisikalo, A.A.Boyarchuk, O.A.Kuznetsov and
V.M.Chechetkin, Astron. Reports {\bf 41}, 786 (1997)
[preprint astro-ph/9802004].

\item
\label{di2} D.V.Bisikalo, A.A.Boyarchuk, O.A.Kuznetsov and
V.M.Chechetkin, Astron. Reports {\bf 41}, 794 (1997)
[preprint astro-ph/9802039].

\item
\label{di3} D.V.Bisikalo, A.A.Boyarchuk, O.A.Kuznetsov and
V.M.Chechetkin, Astron. Reports {\bf 42}, 621 (1998)
[preprint astro-ph/9806013].

\item
\label{di4} D. V. Bisikalo, A. A. Boyarchuk, V. M. Chechetkin,
O.A.Kuznetsov and D.Molteni, Mon. Not. R. Astron. Soc. {\bf
300}, 39 (1998).

\item
\label{di5} D.V.Bisikalo, A.A.Boyarchuk, O.A.Kuznetsov and
V.M.Chechetkin, Astron. Reports {\bf 43}, 229 (1999)
[preprint astro-ph/9812484].

\item
\label{di6} D.V.Bisikalo, A.A.Boyarchuk, O.A.Kuznetsov and
V.M.Chechetkin, Astron. Reports {\bf 43}, 587 (1999)
[preprint astro-ph/0003377].

\item
\label{di7} D.V.Bisikalo, A.A.Boyarchuk, O.A.Kuznetsov and
V.M.Chechetkin, Astron. Reports {\bf 43}, 797 (1999)
[preprint astro-ph/9907084].

\item
\label{di8} D.V.Bisikalo, A.A.Boyarchuk, O.A.Kuznetsov and
V.M.Chechetkin, Astron. Reports {\bf 44}, 26 (2000)
[preprint astro-ph/9907087].

\item
\label{di9} D.Bisikalo, P.Harmanec, A.Boyarchuk and O.Kuznetsov,
Astron. \& Astrophys. {\bf 353}, 1009 (2000).

\item
\label{di10} D.Molteni, D.V.Bisikalo, O.A.Kuznetsov and
A.A.Boyarchuk, Mon. Not. R. Astron. Soc. {\bf 327}, 1103 (2001).

\item
\label{di11} A.A.Boyarchuk, D.V.Bisikalo, O.A.Kuznetsov and
V.M.Chechetkin, \q {\it Mass Transfer in Close Binary Stars}
(Taylor \& Francis, London, 2002).

\item
\label{Tanya98} D.V.Bisikalo, A.A.Boyarchuk, O.A.Kuznetsov,
T.S.Khruzina and \q A.M.Cherepashchuk, Astron. Reports {\bf 42},
33 (1998)
[preprint astro-ph/9802134].

\item
\label{Tanya2000} T.S.Khruzina, A.M.Cherepashchuk, D.V.Bisikalo,
A.A.Boyarchuk and O.A.Kuznetsov, Astron. Reports {\bf 45}, 538 (2001).

\item
\label{Tanya2003a} T.S.Khruzina, A.M.Cherepashchuk, D.V.Bisikalo,
A.A.Boyarchuk and O.A.Kuznetsov, Astron. Reports {\bf 47}, 164
(2003).

\item
\label{Tanya2003b} T.S.Khruzina, A.M.Cherepashchuk, D.V.Bisikalo,
A.A.Boyarchuk and O.A.Kuznetsov, Astron. Reports {\bf 47}, 768
(2003).

\item
\label{ippeg} O.A.Kuznetsov, D.V.Bisikalo, A.A.Boyarchuk,
T.S.Khruzina and \q A.M.Cherepashchuk, Astron. Reports {\bf 45},
872 (2001)
[preprint astro-ph/0105449].

\item
\label{Sawada86} K.Sawada, T.Matsuda and I.Hachisu,
Mon.  Not.  R. Astron. Soc. {\bf 219}, 75 (1986).

\item
\label{spiral1} K.Sawada, T.Matsuda and I.Hachisu, Mon. Not. R.
Astron. Soc. {\bf 221}, 679 (1986).

\item
\label{spiral2} K.Sawada, T.Matsuda, M.Inoue and I.Hachisu,
Mon. Not. R. Astron. Soc. {\bf 224}, 307 (1987).

\item
\label{blob1} D.V.Bisikalo, A.A.Boyarchuk, O.A.Kuznetsov,
A.A.Kilpio and \q V.M.Chechetkin, Astron. Reports {\bf 45}, 611
(2001)
[preprint \q astro-ph/0102241].

\item
\label{blob2} D.V.Bisikalo, A.A.Boyarchuk, O.A.Kuznetsov and
A.A.Kilpio, Astron. Reports {\bf 45}, 676 (2001)
[preprint astro-ph/0105403].

\item
\label{fridman} A.M.Fridman, A.A.Boyarchuk, D.V.Bisikalo,
O.A.Kuznetsov, \q O.V.Khoruzhii, Yu.M.Torgashin and A.A.Kilpio,
Phys. Lett. A {\bf 317}, 181 (2003).

\item
\label{rujichka} K.Kornet and M.R\'o\.zyczka, Acta Astron.
{\bf 50}, 163 (2000).

\item
\label{kumar} S.Kumar, Mon. Not. R. Astron. Soc. {\bf 223}, 225
(1986).

\item
\label{warner} B.Warner, {\it Cataclysmic Variable Stars}
(Cambridge Univ. Press, Cambridge, 1995).

\item
\label{roe}
P.L.Roe, Ann. Rev. Fluid Mech. {\bf 18}, 337 (1986).

\item
\label{osher} S.R.Chakravarthy and S.Osher, AIAA Pap. N 85-0363
(1985).

\item
\label{prep} P.V.Kaygorodov and O.A.Kuznetsov, preprint of
Keldysh Institute of applied mathematics N 59 (2002) [in
Russian].

\end{enumerate}

\end{document}